\documentclass[12pt,onecolumn,final]{IEEEtran}

\usepackage{graphicx}
\usepackage{subfigure}
\usepackage{url}
\usepackage{amsmath}
\usepackage{amssymb}
\usepackage{bm}
\usepackage{rotating}
\usepackage[usenames,dvipsnames]{color}
\begin{document}

\title{A Nonlinear Differential Equation for Generating Warping Function}
\author{Arman~Kheirati~Roonizi
\thanks{A. Kheirati Roonizi is with Department of Computer Science, Faculty of Science, Fasa University, Fasa, Iran}
}
\maketitle
\begin{abstract}
Given set of functions $y_i(t)$ and $x(t)$ such that $y_i(t) = a_i x\left[h_i(t)\right]$ with $a_i$ being an unknown amplitude with low changes in time (or $\frac{\Delta a_i}{a^2_i} << 1$) and $h_i(t)$ an unknown warping function, the paper shows that $h_i(t)$ can be described using a non-linear differential equation.
The differential equation then can be utilized to estimate the warping function $h_i(t)$ using a nonlinear least-squares optimization. 
This differential equation can also be useful for reducing and analyzing phase variability in data sequences. 
Results, obtained on synthetic curves, showed that the proposed method is effective in aligning the curves. The obtained aligned curves exhibit variation only in amplitude, and phase variation can be removed efficiently.
\end{abstract}

\section{Introduction}
\label{sec:introduction}
\begin{figure}
\centering
\vspace{6pc}
\includegraphics[width=.7\columnwidth]{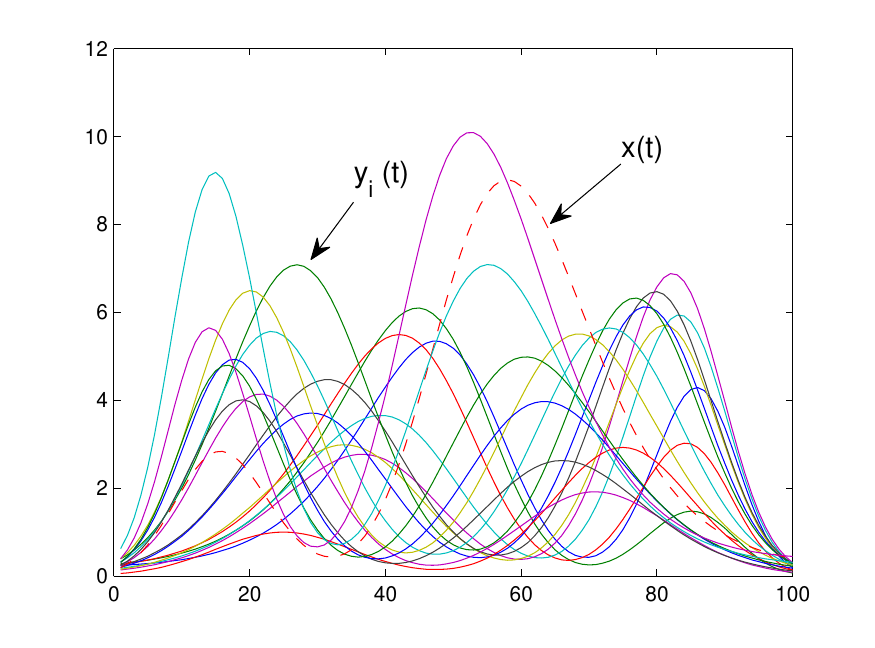}
\caption{21 random unregistered curves with a two-dimensional structure generated from curves (\ref{eq:ysimulation1}) and warping functions (F1)}
\label{fig:variationcurves}
\end{figure}

Curve registration is one of the significant problems in functional data analysis.
For an overview, consider the set of curves shown in Fig(\ref{fig:variationcurves}). These curves differ from each other on the grounds of heights and locations of their peaks and valleys. So we can distinguish the amplitude variability from phase variability by defining the former as associated with the height and the latter with the location of peaks and valleys. Due to these problems, the issue of curve registration has been approached differently by authors.
A process of finding the monotone transformation to align the features of sample of one curve with another is named curve registration in literature review. It has been one of the most challenging problems during the last two decades \cite{Gasser, Kneip2008, Ramsay1998, Wang1997}. 
For more details see the monograph by Ramsay and Silverman \cite{Ramsay2005}.  

Many developments in curve alignment (curve registration) have been proposed in the related literature:

Estimating Smooth Monotone Functions \cite{Ramsay1996}, Continous Monotone Registration \cite{Ramsay1998}, Pairwise Curve Synchronization \cite{Tang2008}, Local Regression and Locally Estimate Monotone Transformation \cite{Kneip2000,Kneip2008}, Curve alignment by equating the moments of a given set of curves \cite{James}, Curve alignment by Dynamic Time Warping \cite{Wang1997}, Synchronizing Sample Curves Nonparametrically \cite{Wang}, Functional Convex Synchronization \cite{Liu2004}, Functional Linear Regression \cite{Malfait2003,Crambes2009} and Self Modeling Warping Function which is a curve alignment, based on a semi-parametric model for the warping functions\cite{Kneip1988,Gervini}.

One of the early works in using time warping was done by Sakoe \textit{et al.} to synchronize speech signals\cite{Sakoe}. In their work they used dynamic time warping, which is a technique that warps the two time series nonlinearly in a way that similar events are aligned by minimizing the distance between them. It is better to point out that DTW in contrary to its name uses nothing dynamic in the process.

In this paper, we show that the warping functions can be described by a non-linear differential equation.

The rest of the article is organized as follows:
A differential equation for generating warping function is introduced in section \ref{sec:warpingestimation}. 
Experiments and simulation Studies are presented in section \ref{sec:Experim}. 
There are some general remarks and suggestions for future works in the last section.
\section{A Nonlinear Differential Equation}
\label{sec:warpingestimation}
Let's consider $N+1$ functions $y_i(t)$ $1 \leq i \leq N$ and $x(t)$ be defined on the close real interval $[0,1]$, which the variation in these curves can be involved both phase and amplitude. 
Suppose that $x(t)$ is the reference function and $y_i(t)$ is the function which warped to $x(t)$ with some specific warping function $h_i(t)$. 
Generally the objective is to find the function of $h_i(t)\in[0,1]$, which aligns the two functions of $y_i(t)$ and $x(t)$. In the simplest case the above definition is defined as:
\begin{equation}
y_i(t) = a_i x\left[h_i(t)\right]\    \  i = 1,...,N 
\label{eq:timewarp1}
\end{equation}  
In this paper, we assume $a_i$ being an unknown with low changes in time ($\frac{\Delta a_i}{a^2_i} << 1$).

First a linear combination of basis functions is used for representing the observations $x(t)$ and $y_i(t)$:
\begin{equation}
\hat{x}(t) = \sum_{k=0}^{N-1} p_k \phi_k(t)
\label{eq:Cdecomposition}
\end{equation} 
\begin{equation}
\hat{y_i}(t) = \sum_{k=0}^{N-1} q_k \phi_k(t)
\label{eq:Ddecomposition}
\end{equation}
where 
\begin{list}{-}{ }
\item $N$ is the number of basis functions in the expansion.
\item $\{\phi_k(t)\}_{k=0}^{N-1}$ should be called the basis of the expansion.
\item $\{p_k, q_k\}_{k=0}^{N-1}$ are the set of corresponding coefficients of the expansions, which depend on $x(t)$ and $y_i(t)$ respectively.
\item $\hat{x}(t)$ and $\hat{y}_i(t)$ design an ``acceptable'' model for $x(t)$ and $y_i(t)$ respectively.
\end{list}
A very important property for an ``acceptable'' model is its ability in function approximation, i.e.,
the model error $e(t) = x(t) - \hat{x}(t)$ should be within an acceptable range. Considering the fact that the observation $x(t)$ might be rather noisy, an ideal model does not necessarily have a zero error. In fact, while the model $\hat{x}(t)$ should overall resemble $x(t)$, there are always some noisy fluctuations within $x(t)$ that should be neglected by the model. In other words, denoising is somewhat intrinsic to modeling. Nevertheless, the basis functions should generally have the property that the energy of approximation error converges to zero as the model order increases ($N\rightarrow \infty$). This property is guaranteed for $\{\phi_k(t)\}$ that form an orthogonal basis (such as sinusoidal basis).

According to what was said, we can have the following definition with a good approximation.
\begin{equation} 
\hat{y}_i(t) \equiv a_i \hat{x}\left[h_i(t)\right] 
\nonumber
\end{equation} 
and
\begin{equation} 
\sum_{k=0}^{N-1} q_k \phi_k(t) =  a_i \sum_{k=0}^{N-1} p_k \phi_k\left[h_i(t)\right]
\label{eq:timewarpmodel}
\end{equation}
Taking the derivative of both sides of (\ref{eq:timewarpmodel}) yields:
\begin{equation} 
\sum_{k=0}^{N-1} q_k \psi_k(t) = \displaystyle \frac{d}{dt} a_i + a_i \displaystyle \frac{d}{dt}h_i(t) \sum_{k=0}^{N-1} p_k \psi_k\left[h_i(t)\right]
\label{eq:LogWarpingFunction}
\end{equation} 
where $\psi_k(t) = \displaystyle \frac{d}{dt} \phi_k(t), k = 0,...,N-1$.

From (\ref{eq:timewarpmodel} and \ref{eq:LogWarpingFunction}) the following equation is achived:
\begin{equation}
\displaystyle \frac{\sum_{k=0}^{N-1} q_k \psi_k(t)}{\sum_{k=0}^{N-1} q_k \phi_k(t)} = \displaystyle \frac{\displaystyle \frac{d}{dt} a_i}{a_i} + \displaystyle \frac{d}{dt}h_i(t)\displaystyle \frac{\sum_{k=0}^{N-1} p_k \psi_k\left[h_i(t)\right]}{\sum_{k=0}^{N-1} p_k \phi_k\left[h_i(t)\right]}\\
\label{eq:dynamicequalation0}
\end{equation}
Assuming $\displaystyle \frac{\displaystyle \frac{d}{dt} a_i}{a_i} << 1$, then eq.~(\ref{eq:dynamicequalation0}) can be written as: 
\begin{equation}
\displaystyle \frac{\sum_{k=0}^{N-1} q_k \psi_k(t)}{\sum_{k=0}^{N-1} q_k \phi_k(t)} = \displaystyle \frac{d}{dt}h_i(t)\displaystyle \frac{\sum_{k=0}^{N-1} p_k \psi_k\left[h_i(t)\right]}{\sum_{k=0}^{N-1} p_k \phi_k\left[h_i(t)\right]}\\
\label{eq:dynamicequalation1}
\end{equation}
which is nonlinear with respect to $h_i(t)$, but no longer depends on $a_i$.

On the other hand we can represent (\ref{eq:dynamicequalation1}) in matrix notation:
\begin{equation}
\displaystyle \frac{\displaystyle   q^T \Psi(t)}{\displaystyle   q^T \Phi(t) } = \displaystyle \frac{d}{dt}h_i(t) \displaystyle \frac{\displaystyle   p^T \Psi(h_i(t))}{  p^T \Phi\left[h_i(t)\right]}\\
\label{eq:dynamicequalation2}
\end{equation}
where
\begin{equation}
\begin{array}{l}
\displaystyle  {\displaystyle   p} = [p_0, p_1, ..., p_{N-1}]^T\\
\displaystyle  {\displaystyle   q} = [q_0, q_1, ..., q_{N-1}]^T\\
\Phi(t) = [\phi_0(t), \phi_1(t), ..., \phi_{N-1}(t)]^T\\
\Psi(t) = \displaystyle \frac{d}{dt} \Phi(t)
\end{array}
\label{eq:coefs}
\end{equation}
After simplifying we get the following non-linear differential equation
\begin{equation}
\displaystyle \frac{d}{dt}h_i(t) = \displaystyle \frac{\displaystyle   q^T \Psi(t)  p^T \Phi\left[h_i(t)\right]}{\displaystyle   p^T \Psi\left[h_i(t)\right]\displaystyle   q^T \Phi(t) }\\
\label{eq:hrecursionequalation}
\end{equation}
The warping function $h_i(t)$ is considered as a strictly monotone function which has a strictly positive first derivative. Built upon the idea of Ramsay, a smooth monotone function $h_i(t)$ can be defined by expressing its derivative, $\frac{d}{dt} h_i(t)$, as the exponential of an unconstrained function $W$ of the form 
\begin{equation}
\frac{d}{dt} h_i(t) = \exp[W]
\label{eq:derivativemonotonefunction}
\end{equation}
This idea was originally presented by Ramsay, in his landmark paper \cite{Ramsay98}, which has ever since been used for various monotone spline modeling \cite[Ch. 6]{Ramsay2005}.

Unconstrained function $W$ can also be considered as a linear combination of B-spline basis functions:
\begin{equation}
W = c^T B(t)
\label{eq:linearhomogen}
\end{equation}
where $B(t) = [B_0(t),B_1(t)...,B_m(t)]$ is the set of basis functions and $  c = c_0, c_1,...,c_m$ is the unknown corresponding coefficients.
The B-spline basis functions consist of polynomial pieces that are smoothly connected together.

So the warping functions are of the form  
\begin{equation}
h_i(t) = \beta_0 + \beta_1 \int{ \exp[c^T B(t)] dt}
\label{eq:monotonefunction}
\end{equation}
Also the derivative of (\ref{eq:monotonefunction}) can be expressed as:
\begin{equation}
\displaystyle \frac{d}{dt}h_i(t) = \exp[c^T B(t)]
\label{eq:derivativemonotonefunction2}
\end{equation}
By substituting (\ref{eq:monotonefunction}) and (\ref{eq:derivativemonotonefunction2}) in (\ref{eq:dynamicequalation2}) the fitting criterion is defined as:
\begin{equation}
\displaystyle \frac{\displaystyle q^T \Psi(t)}{\displaystyle q^T \Phi(t) } =   \exp[c^T B(t)] \displaystyle \frac{\displaystyle   p^T \Psi \left(\beta_0 + \beta_1 \int{ \exp[c^T B(t)] dt}\right)}{  p^T \Phi \left(\beta_0 + \beta_1 \int{ \exp[c^T B(t)] dt}\right)}
\label{eq:measure2}
\end{equation}  
So the error function is defined as: 
\begin{equation}
\rho(t) = \displaystyle \frac{\displaystyle   q^T \Psi(t)}{\displaystyle   q^T \Phi(t) } -  \exp[c^T B(t)] \displaystyle \frac{\displaystyle   p^T \Psi \left(  \beta_0 + \beta_1 \int{ \exp[c^T B(t)] dt} \right)}{  p^T \Phi \left(\beta_0 + \beta_1 \int{ \exp[c^T B(t)] dt} \right)}
\label{eq:measure3}
\end{equation}
where $  p$, $  q$, $\Phi(t)$, $\Psi(t)$ and $B(t)$ previously determined and only unknown parameters is the coefficients vector of $  c$. 
The unknown coefficients $  c$ are evaluated to minimize the corresponding LS criterion:
\begin{equation}
\sigma^2 = \int_{0}^{1}{\left[\displaystyle \frac{\displaystyle   q^T \Psi(t)}{\displaystyle   q^T \Phi(t) } -   \exp[c^T B(t)] \displaystyle \frac{\displaystyle   p^T \Psi \left(  \beta_0 + \beta_1 \int{ \exp[c^T B(t)] dt} \right)}{  p^T \Phi \left(  \beta_0 + \beta_1 \int{ \exp[c^T B(t)] dt} \right)}\right]^2 dt} + \lambda \int_{0}^{1}{\left(1 - \exp[c^T B(t)]\right)^2} dt
\label{eq:measure}
\end{equation}  

Thus the numerical solution of minimizing Eq(\ref{eq:measure}) estimates the warping functions.

For this the criterion (\ref{eq:measure}) must be minimized with respect to the coefficient vector $  c$.
The Matlab function $nlinfit.m$ or $lsqnonlin.m$ performs the required implementation of this nonlinear least-squares optimization.
\begin{figure*}
\centering
\subfigure[the true time transformations $h_i(t)$ (F1)]{\includegraphics[width=.3\columnwidth]{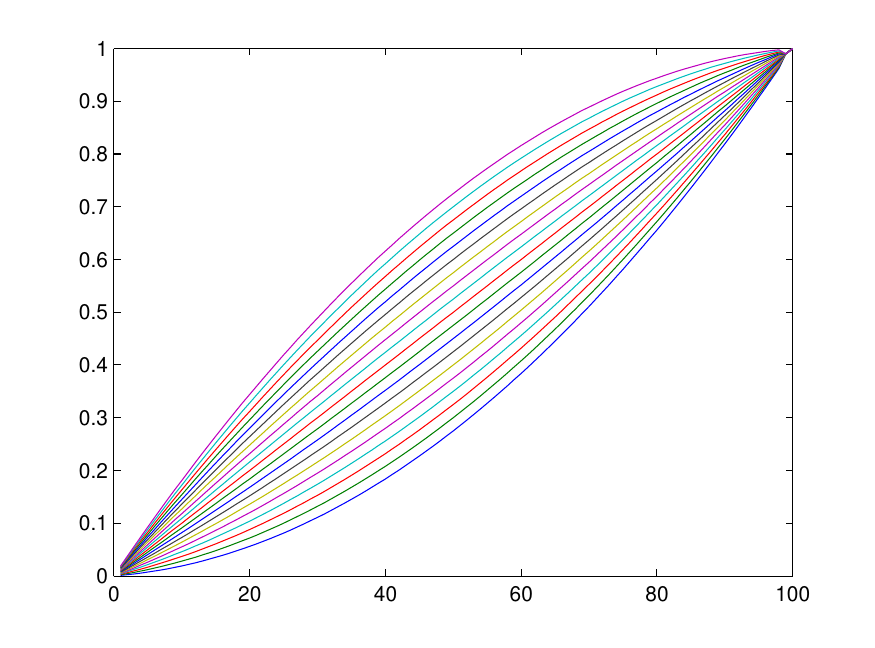}\label{fig:realwrappingpathF1}}
\subfigure[the 21 sample curves before alignment;]{\includegraphics[width=.3\columnwidth]{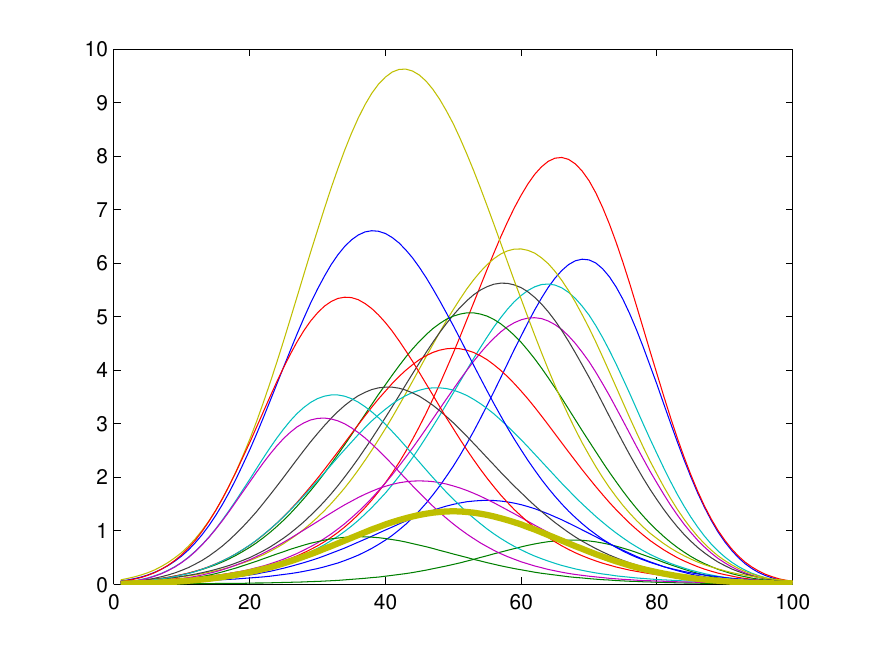}\label{fig:beforealignment1_F1}}	
\subfigure[the curves after alignment]{\includegraphics[width=.3\columnwidth]{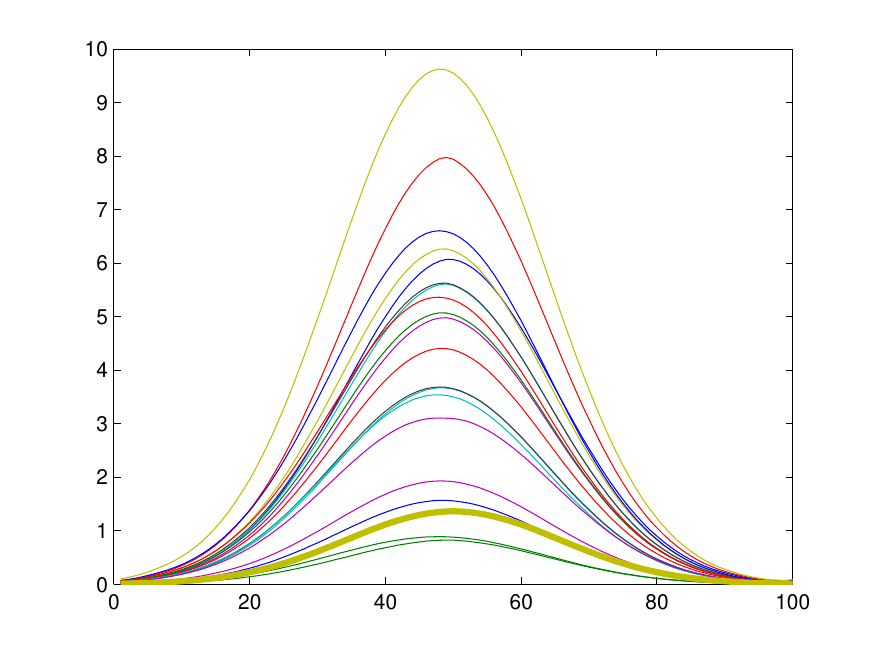}\label{fig:afteralignment1_F1}}
\subfigure[the 21 sample curves before alignment;]{\includegraphics[width=.3\columnwidth]{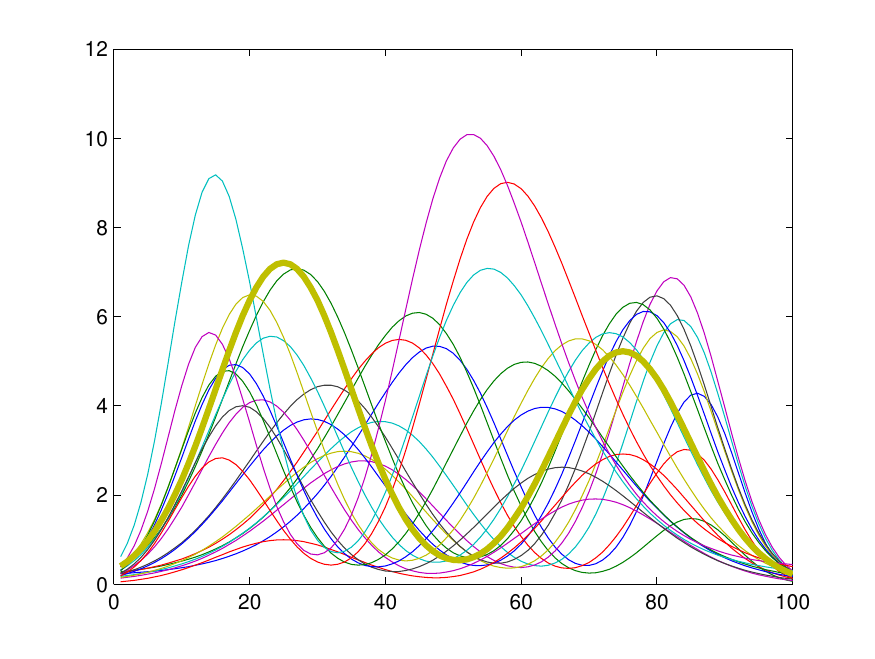}\label{fig:beforealignment2_F1}}	
\subfigure[the curves after alignment]{\includegraphics[width=.3\columnwidth]{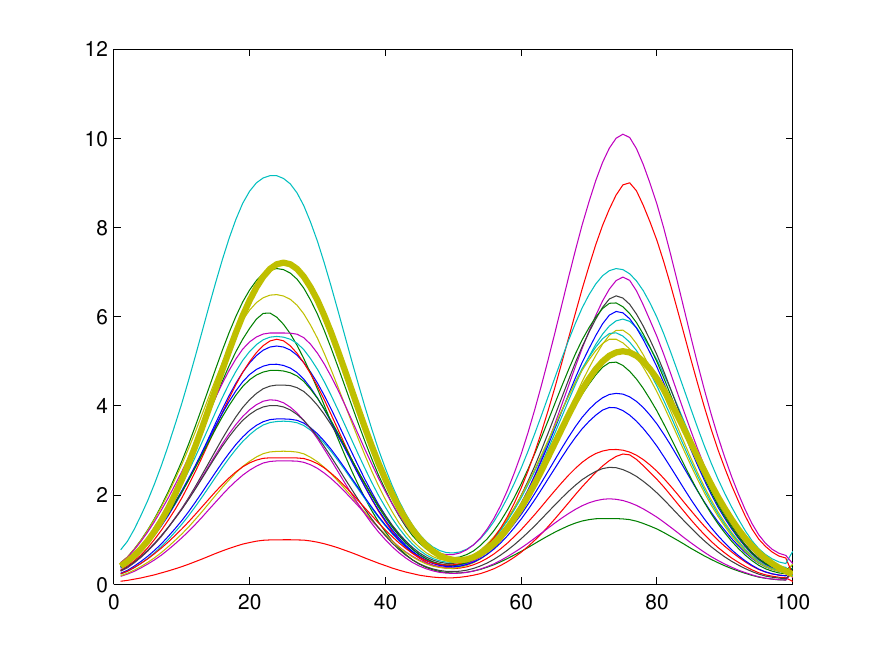}\label{fig:afteralignment2_F1}}
\subfigure[the plot of the functions estimated warping functions $\hat{h}_i(h_i(t))$, which should be $\hat{h}_i(h_i(t)) = t$ for all $i$;]{\includegraphics[width=.3\columnwidth]{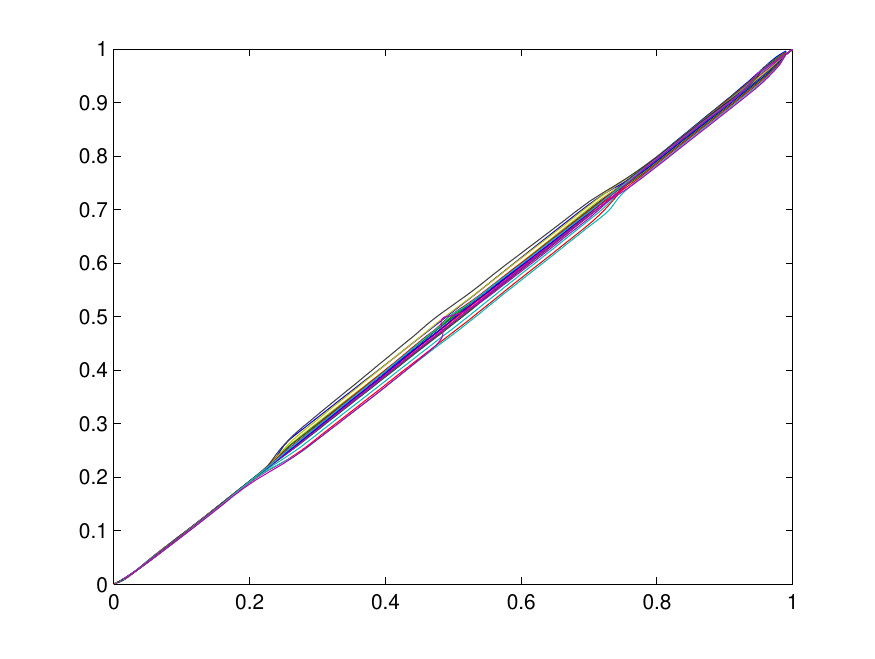}\label{fig:estimatewarpingpath1_F1}}
\caption{A typical run of the simulations with model (F1) for $h_i(t)$ and $z_i$ with distribution $N(5, 1.25)$ for $x(h_i(t))$, Reference function (blue dashed line), average curve(Orange solid line)}.
\label{fig:simul1}
\end{figure*}	

\section{A Simulated Data Illustration}
\label{sec:Experim}
The proposed algorithm was implemented in Matlab. 
To study the performance of the proposed method several simulation data sets were used.
We generated four sets of 21 curves over the interval $[0,1]$, which exhibit variation in amplitude and phase simultaneously:
\begin{equation}
\begin{array}{lll}
\displaystyle y_i(t) = \sum_{k=0}^{N-1} z_{ik} \exp[\displaystyle -\frac{(t - t_k)^2}{2b_k^2}]
\end{array}
\label{eq:ysimulation1}
\end{equation}
where $N$ is the number of Gaussian functions, $b_k$ and $t_k$ are width, and center parameters of the Gaussian terms and the expansion coefficients $z_{ik}$ were randomly generated from the distribution $N(5, 1.5)$. 

These functions are suitable for describing many processes in mathematics, science, and engineering. That is why we used these functions for simulates data. Some typical values of these parameters are listed in Table \ref{tab:typparam}.

\begin{figure*}
\centering
\subfigure[the true time transformations $h_i(t)$ (F2)]{\includegraphics[width=.3\columnwidth]{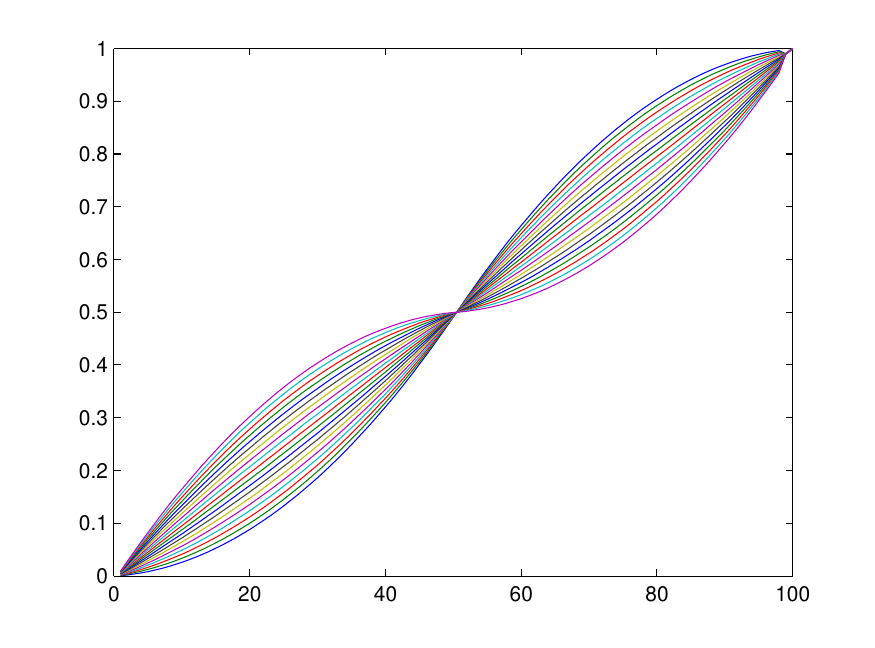}\label{fig:realwrappingpathF2}}
\subfigure[the 21 sample curves before alignment;]{\includegraphics[width=.3\columnwidth]{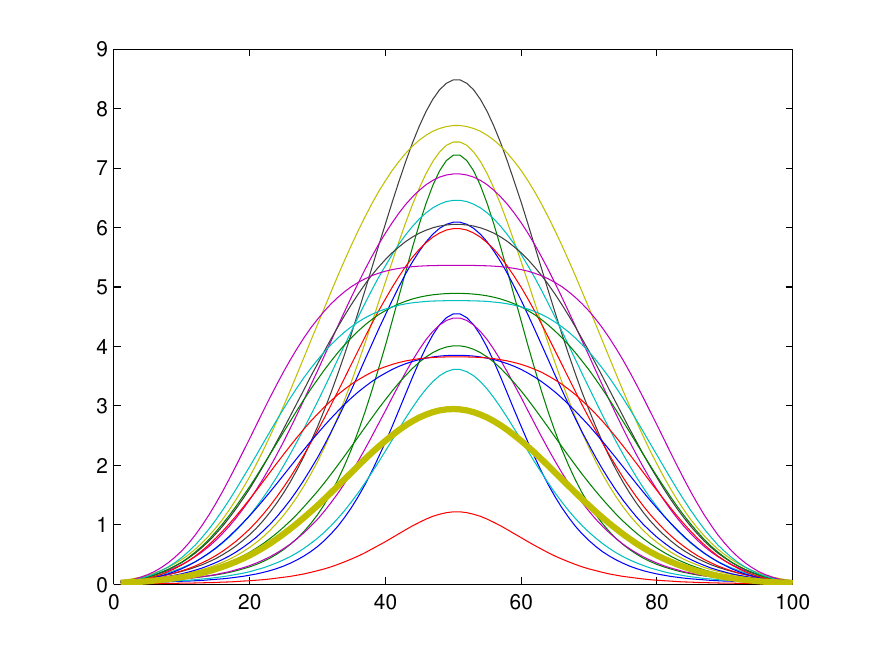}\label{fig:beforealignment1_F2}}	
\subfigure[the curves after alignment]{\includegraphics[width=.3\columnwidth]{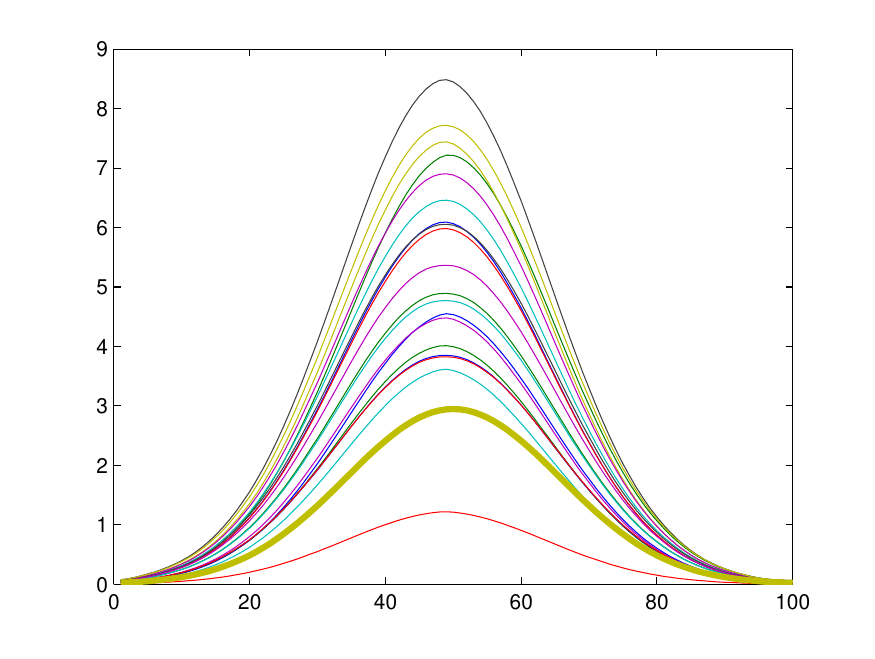}\label{fig:afteralignment1_F2}}
\subfigure[the 21 sample curves before alignment;]{\includegraphics[width=.3\columnwidth]{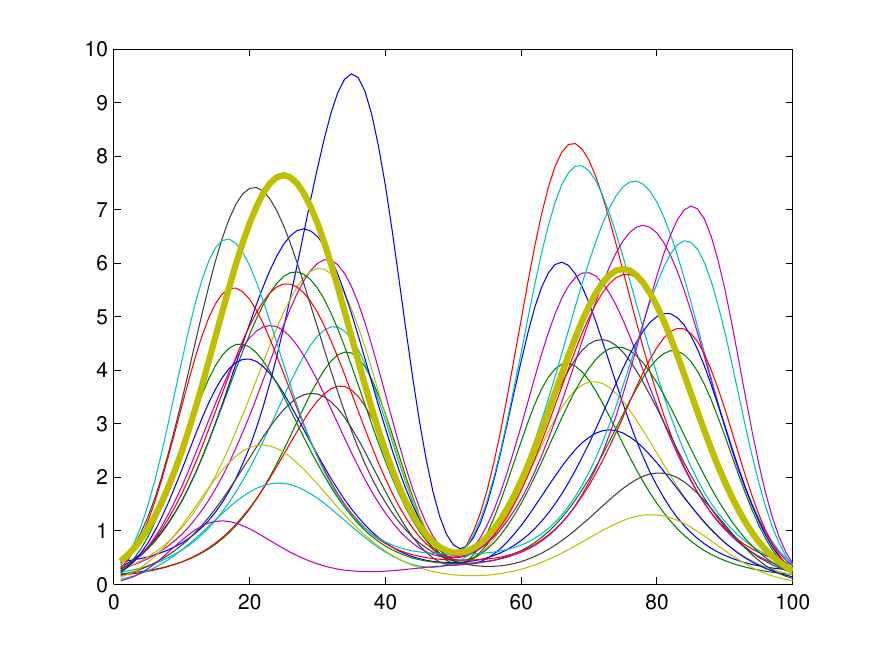}\label{fig:beforealignment2_F2}}	
\subfigure[the curves after alignment]{\includegraphics[width=.3\columnwidth]{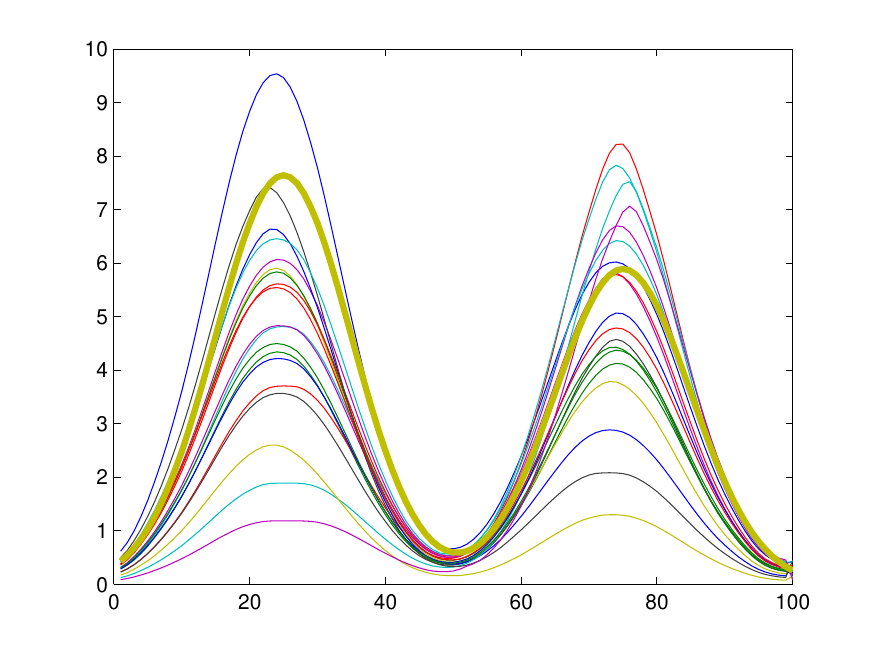}\label{fig:afteralignment2_F2}}
\subfigure[the plot of the functions estimated warping functions $\hat{h}_i(h_i(t))$, which should be $\hat{h}_i(h_i(t)) = t$ for all $i$;]{\includegraphics[width=.3\columnwidth]{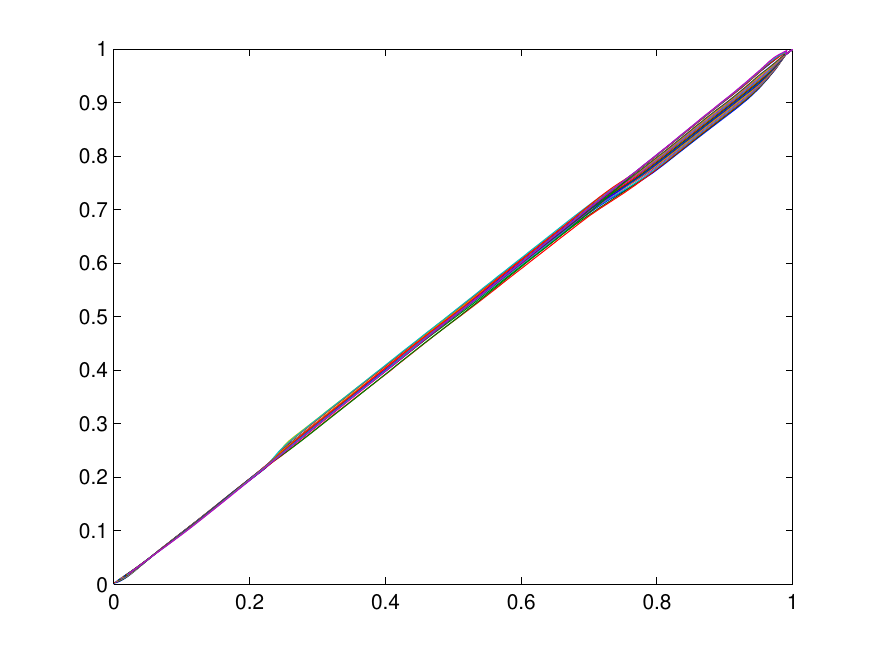}\label{fig:estimatewarpingpath1_F2}}
\caption{A typical run of the simulations with model (F2) for $h_i(t)$ and $z_i$ with distribution $N(5, 1.25)$ for $x(h_i(t))$, Reference function (blue dashed line), average curve(Orange solid line)}.
\label{fig:simul3}
\end{figure*}

We consider two forms of the time transformations $h_i(t)$:
\begin{enumerate}
\item \textbf{F1:} Quadratic transformations $h_i(t) = t + b_it(1 - t)$, and the coefficients $b_i$ were equally spaced between $-1$ and $1$.
\item \textbf{F2:} $h_i(t) = t + b_isin(2 \pi c_it)$ where $c_i \in \{0,1,2,3\}$. For more details see \cite{Wang}.
\end{enumerate}
Different examples of the curves $y_i(t)$ are shown in figures \ref{fig:simul1} and \ref{fig:simul3}. 

Before aligning the curves, we need to choose the reference curve ($x(t)$) from the existing curves $y_i(t)$. The algorithm is described in Appendix. 
   
The basis functions $\Phi(t)$ which were used to model the unregistered curves $y_i(t)$ consisted of Sinusoidal or B-spline basis functions. In this study we employed the sinusoidal basis.

The standard quantitative measurement is the percentage root-mean-square difference (PRD), which is given by:
\begin{equation}
PRD \equiv 100\sqrt{\displaystyle \frac{\int_{-\infty}^{\infty}{(\hat{y}_i(t) - \hat{x}(\hat{h}_i(t)))^2}dt}{\int_{-\infty}^{\infty}{{\hat{y}_i^2(t)}dt}}}
\label{eq:PRD}
\end{equation} 
where, $h_i$ and $\hat{h_i}$ are original and estimated functions. We used PRD for comparing the performance of the order of the model, where a lower PRD value indicates that the reconstruction approximates the original
more closely and is therefore better. The results of comparing the performance of Sinusoidal and B-spline basis functions which are defined by equally spaced knots, are summarized in table \ref{tab:quantitativecomp}.
Considering that the original curve has 1000 samples, the number of coefficients in \ref{eq:linearhomogen} vary between ($N = 10 : 5 : 45$). It is obvious that the value of PRD decreases while the order of the model increases. 


\begin{table*}
\centering\footnotesize
\caption{Parameters of simulated model in (\ref{eq:ysimulation1})}
\begin{tabular}{llll}
Parameters &  N = 1 &  N = 2 \\
\hline
$t_k$ & 0.5 & [0.25 0.75]  \\
$b_k$ & 0.1581 & [0.1 0.1] \\
\end{tabular}
\label{tab:typparam}
\end{table*}
\begin{table*}
\centering\footnotesize
\caption{The results of Sinusoidal and Bspline Models with Percentage root mean square differences, for different order of model for simulated curves(\ref{eq:ysimulation1}) with 1000 samples}
\begin{tabular}{lllllllll}
Method &  Order = 10 & Order = 15 & Order = 20 & Order = 30 & Order = 35 & Order = 40 & Order = 45 \\
\hline
B-Spline &  2.38\% & 2.22\% & 1.33\% & 1.013\% & 0.97\% & 0.83\% & 0.79\% \\
Sinusoidal &  2.87\% & 1.99\% & 1.32\% & 0.98\% & 0.88\% & 0.84\% & 0.78\% \\
\end{tabular}
\label{tab:quantitativecomp}
\end{table*}
\section{Discussion and Conclusion}
\label{sec:conclusion}
In this paper, a nonlinear differential equation was introduced for describing warping functions. 
The proposed equation was also used for estimating warping functions. The proposed method is useful for estimating warping function even if the domain of the reference function and the function to be warped are different. 
Unlike Dynamic Time Warping (DTW), the proposed technique has dynamic behavior, so it can be combined within Kalman structure to improve the estimation of warping functions.
In this paper, based on estimating warping function which occurs between two one-dimensional curves, it is easy to extend its principles to high-dimensional curves.  
\appendix
If the reference function, $x(t)$, is unknown then it can be selected from the existing functions $y_i(t)$.
Lets assume that $x(t) = y_j(t)$ then the problem changes to $y_i(t) = a_i y_j\left[h_i(t)\right]$.

After estimating $h_i(t)$ using the algorithm proposed in this paper, the amplitude $a_i$ can be found by minimizing $\int_{0}^{1} \left[y_i(t) - a_i y_j(h_i)\right]^2$.
Also we have
\begin{equation}
h_i(t) = y_j^{-1} \left[\frac{y_i(t)}{a_i}\right]\    \  i = 1,...,N 
\label{eq:p1}
\end{equation}  
We define the following criterion: 
\begin{equation}
J_j = \frac{1}{N} \sum_{i=1}^{N} \left[h_i(t) - t\right]^2 = \frac{1}{N} \sum_{i=1}^{N} \left[y_j^{-1} \left(\frac{y_i}{a_i}\right) - t\right]^2
\label{eq:p2}
\end{equation} 
We compute the following criterion changing the reference function $y_i(t) = a_i y_j\left[h_i(t)\right]$, $i,j = 1,...,N $: 
which is the power of the error between the estimated warping functions and $h_i(t) = t$. Actually we are trying to find the best reference which satisfy the regularization part in eq.~(\ref{eq:measure}).

It will be satisfied if $y_j$ minimizes criterion (\ref{eq:p2}).

Another possibility is to use the following algorithm:

Considering that the reference function $x(t)$ can shifted by warping functions $h_i(t)$. Therefore the energy of the function can also be shifted.
 
first we computed the energy of the observed functions but for half time
\begin{equation}
Power_i = \int_{0}^{\frac{1}{2}} y^2_i(t) dt\    \  i = 1,...,N 
\label{eq:p3}
\end{equation}
Then we sort the Power of the signals.
The curve which its power is equal to median(Pow) can be selected as a reference function.
\bibliographystyle{IEEEtran}
\bibliography{References}
\end{document}